# The Design and Performance of Charged Particle Detector onboard the GECAM Mission

Y.B. Xu; X.Q. Li; X.L. Sun; S. Yang; H. Wang; W.X. Peng; X.H. Liang; K. Gong; Y.Q. Liu; D.Y. Guo; X.Y. Zhao; C.Y. Li; Z.H.An; J.J. He; X.J. Liu; X.Y. Wen; S.L. Xiong; Fan Zhang; D.L. Zhang; C.Y. Zhang; C. Cai; Z. Chang; G. Chen; C. Chen; Y.Y. Du; M. Gao; R. Gao; D.J. Hou; Y.G. Li; G. Li; L. Li; X.F. Li; M.S. Li; F.J. Lu; H. Lu; B. Meng; F. Shi; J.Z. Wang; Y.S. Wang; H.Z. Wang; X. Wen; S. Xiao; Y.P. Xu; J.W. Yang; Q.B. Yi; S.N. Zhang; C.M. Zhang; F. Zhang; Y. Zhao; X. Zhou;

(Key Laboratory of Particle Astrophysics, Institute of High Energy Physics, Chinese Academy of Sciences, Beijing, 100049, People's Republic of China)

**Abstract:** The Gravitational Wave highly energetic Electromagnetic Counterpart All-sky Monitor (GECAM) is dedicated to detecting gravitational wave gamma-ray bursts. It is capable of all-sky monitoring over and discovering gamma-ray bursts and new radiation phenomena. GECAM consists of two microsatellites, each equipped with 8 charged particle detectors (CPDs) and 25 gamma-ray detectors (GRDs). The CPD is used to measure charged particles in the space environment, monitor energy and flow intensity changes, and identify between gamma-ray bursts and space charged particle events in conjunction with GRD. CPD uses plastic scintillator as the sensitive material for detection, silicon photomultiplier (SiPM) array as the optically readable device, and the inlaid Am-241 radioactive source as the onboard calibration means. In this paper, we will present the working principle, physical design, functional implementation and preliminary performance test results of the CPD.

# 1 Introduction

The existence of gravitational waves was predicted by Albert Einstein in 1916 based on the general theory of relativity [1]. On September 14, 2015, the Laser Interferometer Gravitational-Wave Observatory (LIGO) discovered the gravitational-wave signal originating from the merger of two black holes for the first time [2]. On August 17, 2017, the Advanced LIGO and Advanced Virgo jointly detected gravitational waves (GW170817) originating from a binary neutron star coalescence, opening the era of multi-signal gravitational-wave astronomy [3-11]. According to existing studies, the merger process of binary compact stars not only produces gravitational waves, but also is often accompanied by X-ray/γ-ray, soft X-ray, optical, radio and other wavelengths of radiation [12]. Among the joint multi-wavelength observations of gravitational waves, the observation of gamma-ray bursts can not only provide a trigger for the observation of other wavelengths, but also provide more precise position information for follow-up observations [13, 14]. Therefore, the detection of gravitational wave gamma-ray bursts becomes considerably critical. The Gravitational Wave highly energetic Electromagnetic Counterpart All-sky Monitor (GECAM) project is an exploration project in space science dedicated to the detection of gravitational wave gamma-ray bursts [15]. Its principal goal is to discover the largest sample of gravitational wave gamma-ray bursts and new radiation phenomena, and study compact objects such as neutron stars and black holes and their merging processes.

GECAM is composed of two microsatellites (~162 kg/pc), each consists of 25 gamma-ray detectors (GRDs) and 8 charged particle detectors (CPDs). GRD and CPD are used to detect gamma-rays and charged particles respectively. The two GECAM satellites operate in the same low earth orbit (∼600 km) but in opposite phase, thus can provide complete coverage of the entire sky [17].

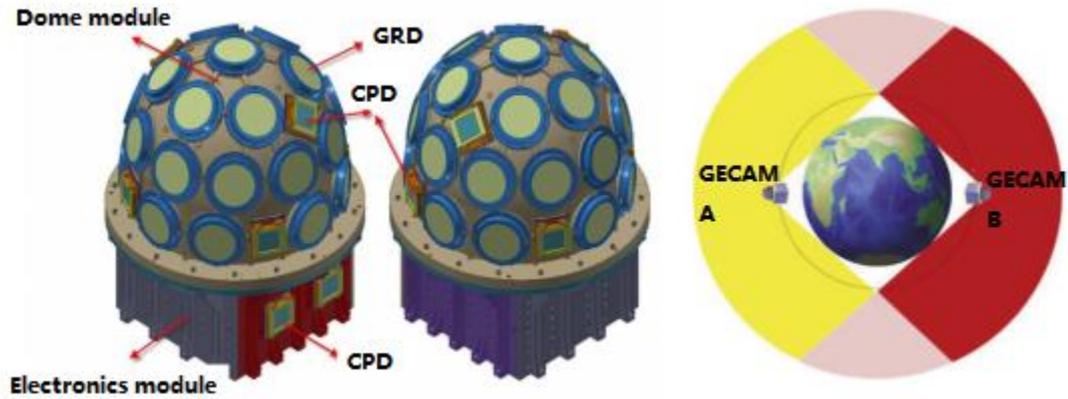

Figure 1 Two GECAM satellites and installation layout of GRDs and CPDs

There are eight CPDs installed in single satellite, including six in the Dome module and two in the electronics module. CPD is mainly used for the measurement of charged particles in the space environment, monitoring changes of their energies and flow intensity. In addition, CPD can distinguish gamma-ray bursts and charged particle events, so as to identify the gravitational wave gamma-ray bursts.

The main components of charged particles in near-Earth space are electrons and protons, which are principally distributed in the radiation belts. They carry a large amount of physical information concerning highly energetic astrophysics, solar physics, interplanetary space physics, and even magnetospheric physics. From the application point of view, they are also involved in radiation biology, component irradiation, etc. Therefore, the detection of charged particles in space is an urgent need in many research fields.

The detectors for energetic charged particle detection in space are roughly divided into two categories: one is magnetic spectrometer, and the other is a calorimeter, or magnetic spectrometer combined with calorimeter. As for the domestic satellites, such as FY series satellites, SZ series satellites, Dark Matter Particle Explorer (DAMPE) and ZH-1 satellite that have carried the particle detectors, most of the sensor parts are semiconductor detectors, mainly for space environment monitoring and particle energy spectrum and flux measurement[19-24]. For the overseas satellites, such as ARINA, SOHO, DEMETER and Van Allen which have carried particle detectors, they mainly use silicon detectors and scintillators to measure energy, and use position-sensitive silicon strip detectors or multiple plastic scintillators to measure the direction of incidence of charged particles and achieve particle identification [25-29]. Different detectors vary in size and specific design as they are used for different principal purposes. As for CPD, plastic scintillators combined with SiPM array were used for space charged particle detection. In this paper, we will present the working principle, physical design and performance testing results of the CPD.

## 2 Working Principle and Structure

CPD mainly applies to detect space electrons at 300 keV-5 MeV. By monitoring the charged particle flow intensity in the space environment, CPD can identify gamma-ray bursts and space charged particle events to achieve the identification of space particle bursts. In addition, CPD can study the onboard background of GECAM. To meet the detection requirement for CPD, a design scheme was adopted where the plastic scintillator is used as a sensitive material for detection and silicon photomultiplier (SiPM) as an optical readable device.

The plastic scintillator has a poor optical response to gamma rays but a good response to charged particles. It also has the advantages of short optical response time, strong irradiation resistance, non-easy deliquescence and easy processing. The BC-408 plastic scintillator was selected for its wavelength characteristics are well matched with SiPM and it has a good optical yield energy linearity for electrons (Figure 2) [30].

The SiPM array was used for light readout. With superior capabilities in photon counting and detection of weak light signals, SiPM has advantages of magnetic field immunity, low bias voltage, high gain, and small size. SiPM is composed of multiple APD arrays operating in Geiger mode, where each APD is a pixel that outputs a charge pulse signal when receiving a photon. The sum of the charges output by all pixels is proportional to the total number of photons detected by the SiPM.

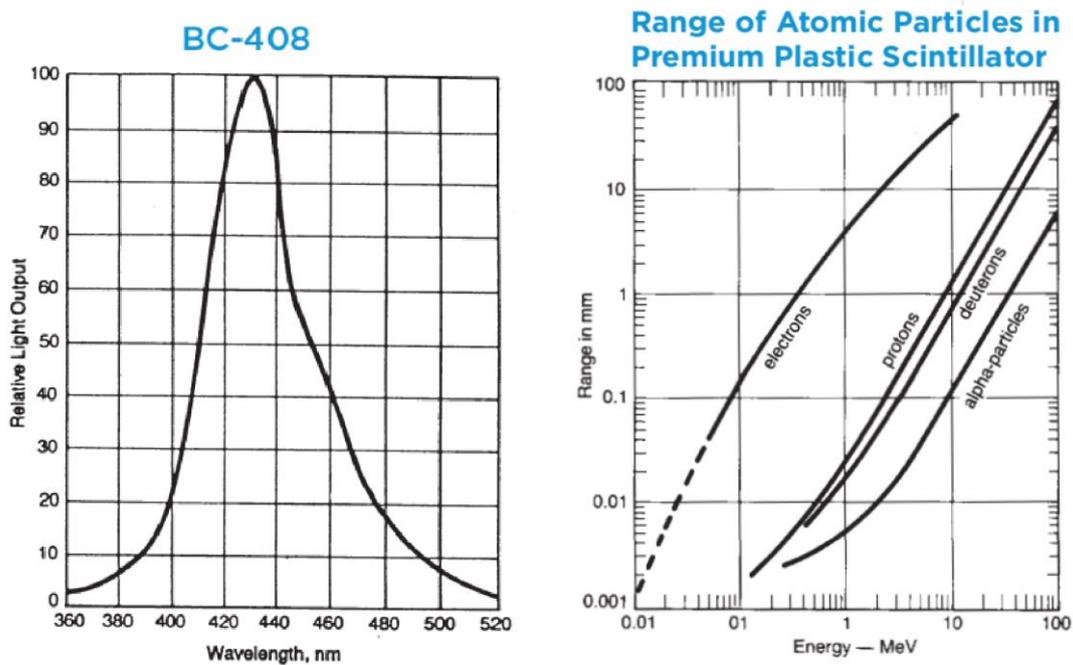

Figure 2 Physical properties of plastic scintillator (BC408)

The actual deposition energy of electrons in the plastic scintillators with different thicknesses was simulated by means of the CPD mass model established in GEANT4 [31-33]. The energy range is from 200 to 5000 keV, and the thickness of plastic scintillators is 6 mm, 10 mm and 20 mm, respectively. The electron deposition energy spectra are shown in Figure 3, where the incident energy, detection efficiency and full energy peak efficiency (FEPE) are shown as labels.

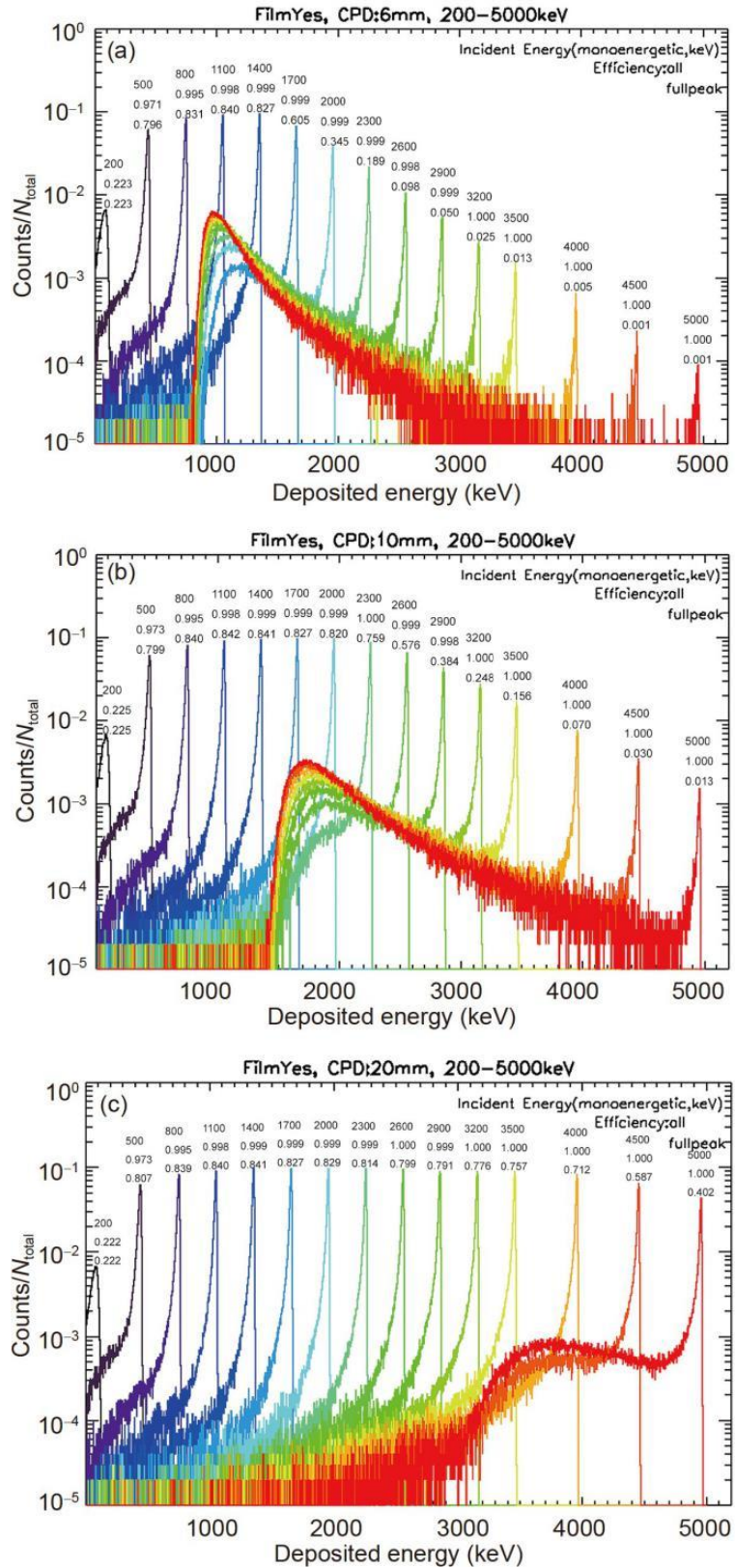

Figure 3: Electron deposition energy spectra of plastic scintillators with three thicknesses. Different color curves correspond to different electron incident energies. (a) 6 mm; (b) 10 mm; (c) 20 mm.

Electrons in near-earth orbit follow a power exponential distribution, and the main components are low-energy electrons. The main detection target of CPD is the electrons below 1 MeV. The energy deposition of highly energetic electrons in the 6 mm plastic scintillator is mostly below 1 MeV, which is not conducive to the energy spectrum detection of the electron energy region below 1 MeV. As a result, the 6 mm-thick solution was excluded first. In contrast, the 20 mm-thick plastic scintillator is better than the 10 mm-thick solution for electron detection, but its thickness is too large and structurally unfavorable for assembly, which is a disadvantage for microsatellite structures. For the 10 mm-thick plastic scintillator, although most of the highly energetic band electrons above 2 MeV are penetrating instances, their energy deposition is above 1.5 MeV. Such electrons therefore have negligible effect on the energy spectrum measurement of the core detection energy region below 1 MeV and does not affect the identification capability of the detector for highly energetic electron bursts. As a result, 10 mm was chosen as the thickness of the CPD plastic scintillator.

By referring to the conventional method of utilizing inlaid radioactive sources to perform onboard calibration of detector [34], the onboard gain calibration was conducted for CPD by inlaying a radioactive source inside the plastic scintillator of each CPD. The inlaid Am-241 α source had an activity of ~20 Bq. The source was electroplated on the surface of a aluminium cylindrical with a diameter of 2.5 mm and a height of 3 mm. A hole was punched in the center of the plastic scintillator, and the source was inlaid and sealed by a cover made of the same material as the plastic scintillator (see Figure 4). The CPD background spectrum with the inlaid Am-241 α source is shown in Figure 4. The Full Energy Peak was contributed by the α particles emitted from AM-241 source.

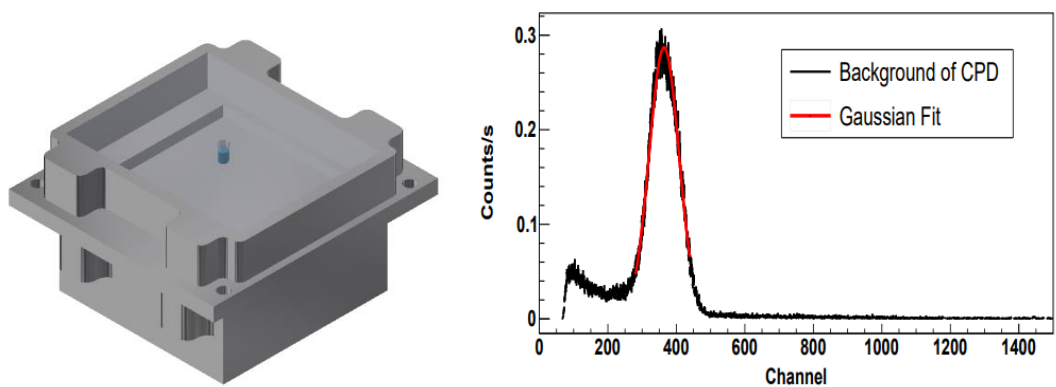

Figure 4 Left: schematic diagram of the CPD inlaid with the radioactive source; Right: background spectrum of the state of CPD probe inlaid with the calibrated source.

As the detection efficiency of gamma rays and charged particles on GRD and CPD differ greatly, the ratio of GRD and CPD count rates can be used to infer whether the burst is composed of gamma rays or charged particles (electron-based). As shown in Figure 5, the CPD has a low detection efficiency only for gamma rays above 200 keV and no response for gamma rays below 200 keV, while it has a high detection efficiency for electrons above 200 keV. According to the difference of detection efficiency in GRD and CPD for gamma rays and electrons, we simulated the response of gamma-ray burst and electron burst on GRD and CPD. The simulation results verified the capability of GRD and CPD in jointly identifying gamma-ray bursts (Figure 6).

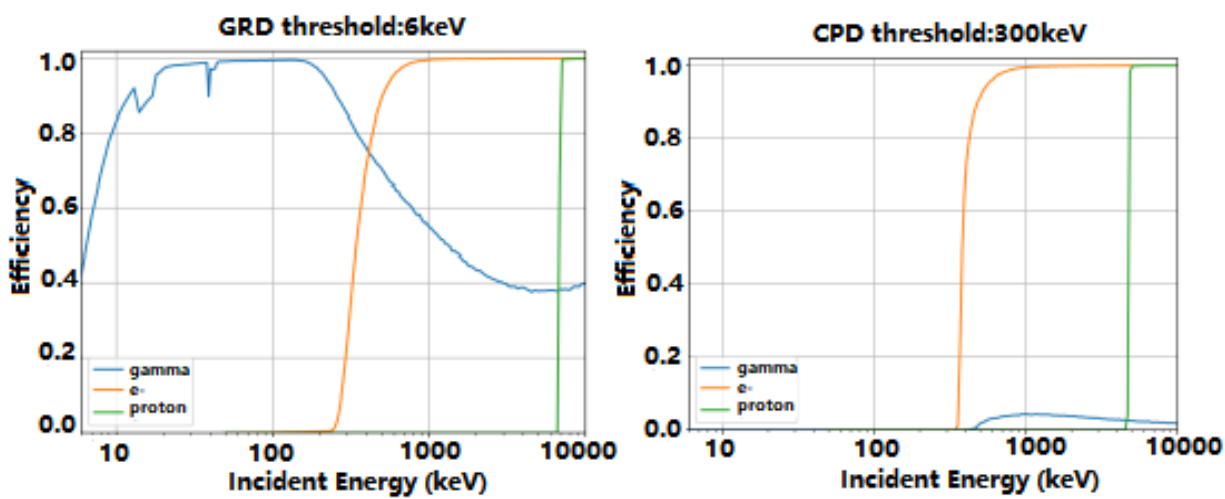

Figure 5 Simulation results of the detection efficiency of GRD and CPD for gamma rays and charged particles

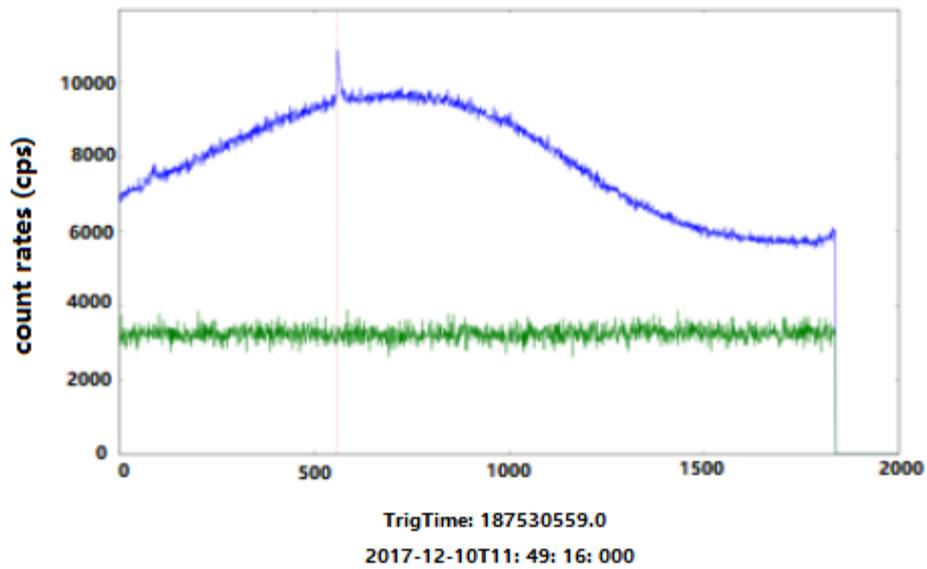

Figure 6 Difference of GRD and CPD in the response of a typical gamma-ray bursts (CPD: green line, GRD: blue line)

The schematic diagram of the CPD structure is shown in Figure 7. It included the upper part of the plastic scintillator box and the lower part of the electronics box, both of which were connected by means of screwing. The structural housing is made of aluminum. The plastic scintillator box was used to mount the plastic scintillator, and all ends of the plastic scintillator were polished and wrapped with Tyvek reflective film to increase light collection. The outer layer of the reflective film was wrapped by shading materials, which shaded the light while playing a role in shock absorption. The outermost layer was an outer protective layer made of polyimide aluminized film to protect the detector from atomic oxygen damage in the space environment.

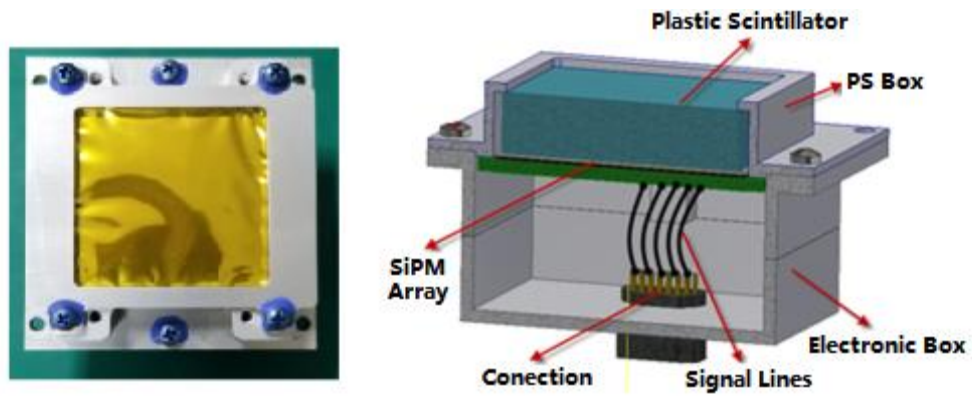

Figure 7 The shape and the structure diagram of CPD

The electronics box consists of a Printed Circuit Board (PCB), a structure and a connector, with the SiPM array on the front side of the PCB and the front-end electronics section on the back side. We used the MicroFJ-60035-TVS SiPM from SensL. The SiPM array was square in shape and consisted of 6x6 SiPM units with a total size of 40 mm*40 mm (Figure 8). Each SiPM unit has a size of 6.07×6.07 mm, with a pixel size of 35 μm. All 36 SiPM units were divided into two groups for power supply, and are combined into one route for readout. The SiPM array and the plastic scintillator were coupled together by a 1 mm-thick optical silicone pad, which not only increased the light harvesting efficiency, but also acted as a shock absorber. An electrical connector was mounted on the outside of the electronics box to electrically connect the CPD to the load processor. Because of the difference in mounting position, the CPD in the dome module and the electronics module differed in connector directions.

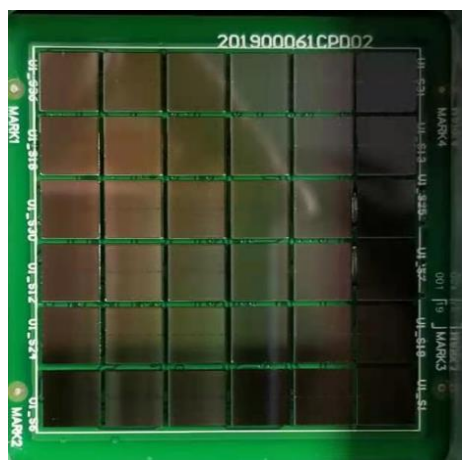

Figure 8 SiPM array on CPD

The front-end electronics part of the CPD consisted of a front-end amplifier circuit with discrete components and a temperature sensor. The SiPM output signal was amplified in two stages by two LM6172 amplifiers and passed as a differential signal to the back-end data acquisition circuit for digitization. The front-end amplifier circuit was connected to the back-end data acquisition circuit through a 120Ω matching resistor. The maximum output voltage of the differential signals was +3.2V and -3.2V respectively, with a conventional signal rise time of 200ns and fall time of 800 ns. DS18B20Z was used to monitor the temperature of SiPMs. The block diagram of the single

CPD circuit is shown in Figure 9.

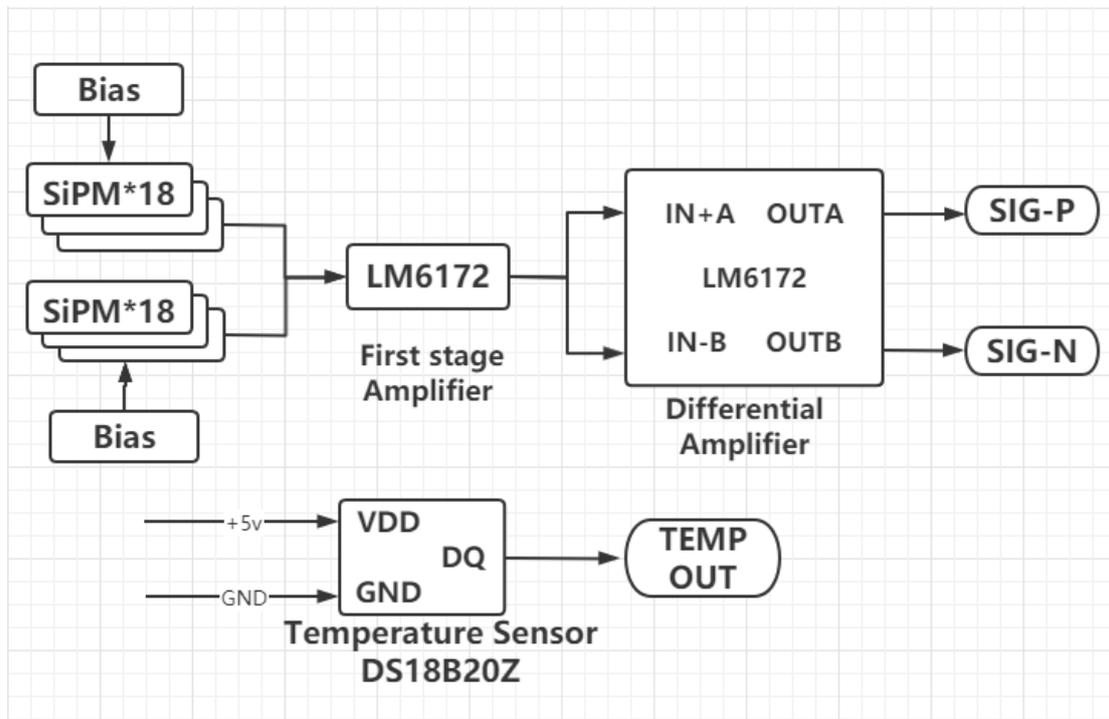

Figure 9 CPD single probe circuit block diagram

As CPD will be positioned on the surface of the satellite, the temperature control unit keeps SiPM working in the range of (-20±3)°C, with certain temperature fluctuation. Since the SiPM had a significant temperature drift characteristic, a temperature change of 30°C would lead to a doubled gain of SiPM signal, so it was necessary to make a temperature correction for the SiPM. The SiPM temperature was collected by the temperature sensor inside the detector and fed back to the power supply circuit for SiPM bias adjustment. Real-time temperature correction of the gain was achieved when the ambient temperature changed, thus maintaining the stability of the SiPM signal gain.

As the SiPMs for each CPDs were divided into two independent groups with independent power supply based on reliability considerations, there were two operating modes for each CPD: full-component mode and semi-component mode.

Under normal conditions CPD operated in semi-component mode. This means that half of SiPM of CPD was powered. In some special cases, such as short circuit or excessive noise of individual SiPM unit, CPD could switch to another semi-component to work. Also CPD could be switched to full-component mode when a higher resolution was required for the measurement target or when fine measurement of electrons was needed in the lower energy range.

# 3 Experiments and Results Analysis

At the Beijing Key Laboratory of Space Environment Exploration，National Space Science Center of the Chinese Academy of Sciences, we used the High Energy Electronics Test System to calibrate the electronic response of CPD in the low-energy region (250keV ~ 1800keV). The results of E-C relationship in full-component mode and semi-component mode are shown in Figure 10 and Figure 11, respectively. It can be seen that the E-C relationship has good linearity in the range of 250-1800 keV for both modes, and the fitting residuals are basically within 1%.

As shown in Fig.10 and figure 11, the channel-energy relation both satisfy good linearity for the Full-component mode and semi-component mode. The SiPM gain of the full-component mode is about twice that of the semi-component mode. In the semi-component mode, CPD has higher detection energy range than the full-component mode. With E-C relationship, the energy range of CPD can be calculated through lower threshold and upper limit of ADC to be 210 keV ~ 6.3 MeV .

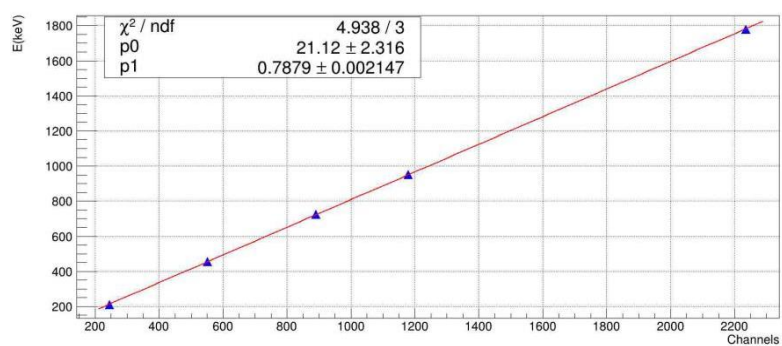

Figure 10: E-C relationship in full-component mode

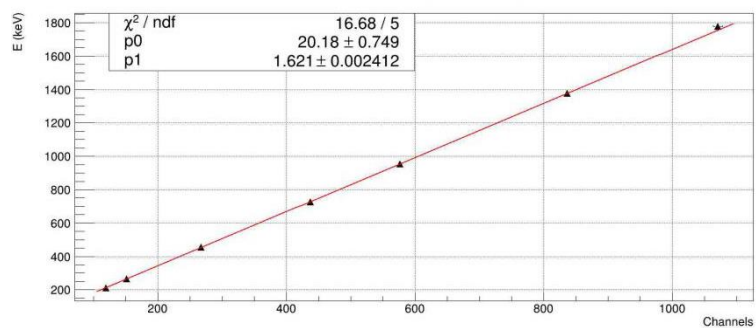

Figure 11 E-C relationship in semi-component mode

To test the identification capability of GRD and CPD for gamma rays, we measured the relative detection efficiency of CPD and GRD for the same Cs-137 (662 keV) gamma-ray source under the same conditions. The relative detection efficiency ratio of CPD and GRD ($Eff_{CPD}/Eff_{GRD}$) could be obtained by counting the over-threshold area in the deposited energy spectrum. According to the results, $Eff_{CPD}/Eff_{GRD}$ is 0.129±0.01. The Cs-137 gamma energy spectra measured by CPD and GRD are shown in Figure 12.

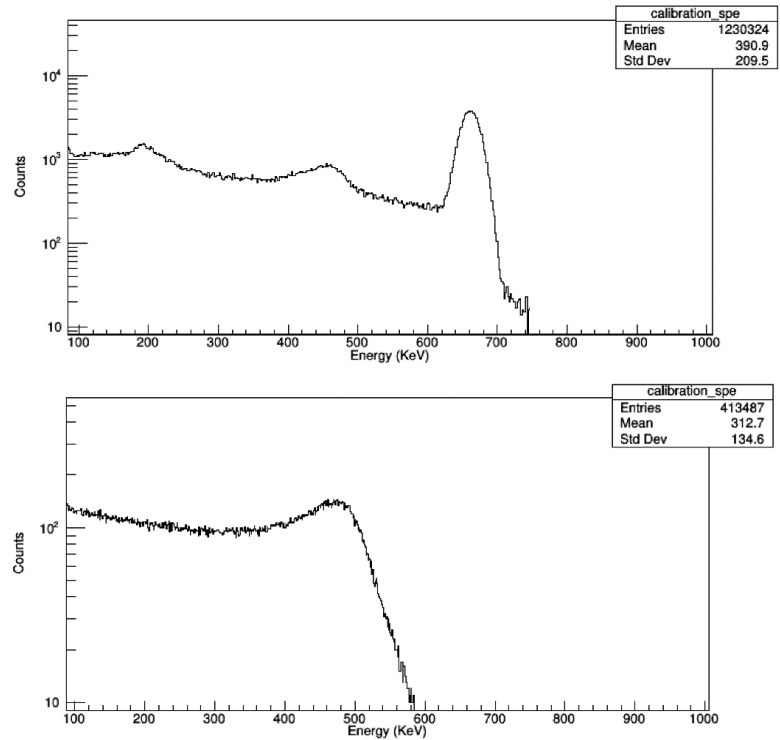

Figure 12 Cs-137 energy spectrum measured (GRD in the upper panel, CPD in the lower panel)

GEANT4 was used to simulate the energy deposited spectrum of Cs-137 gamma-ray source in CPD and GRD detectors, as shown in Figure 13. The $Eff_{CPD}/Eff_{GRD}$ was obtained to be 0.126±0.01 through simulated deposited energy spectrum, which is in good agreement with the experimental results.

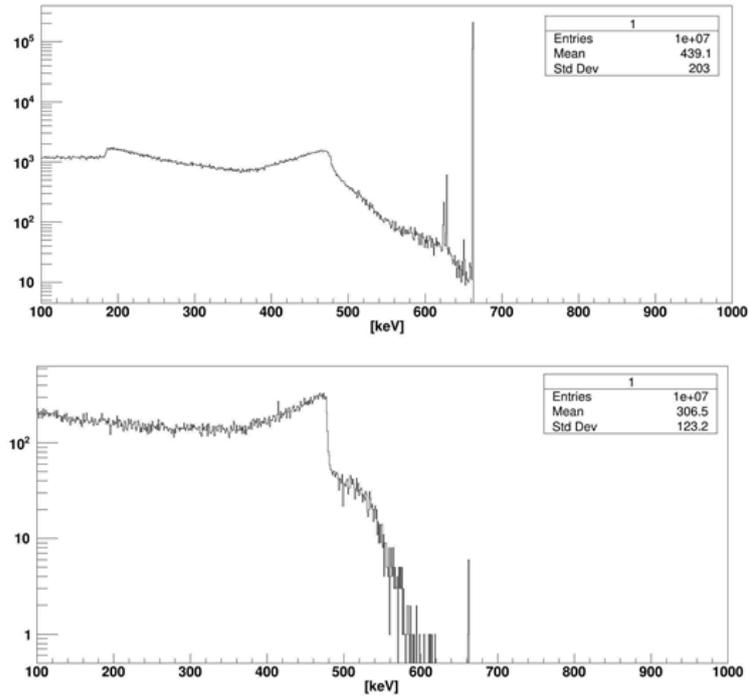

Figure 13 Cs-137 gamma simulated energy spectrum (GRD in the upper panel, CPD in the lower panel)

In order to get the dead-time of the detection system, the high count rate measurement was performed by turning down the threshold value. Since the achieved experimental data containing the time information of each triggered events, the dead time of the system could be obtained from the arrival time interval spectrum. Figure 14 shows the time interval spectrum measured by CPD, and the result shows that the dead time of the whole detection system is 4.8us.

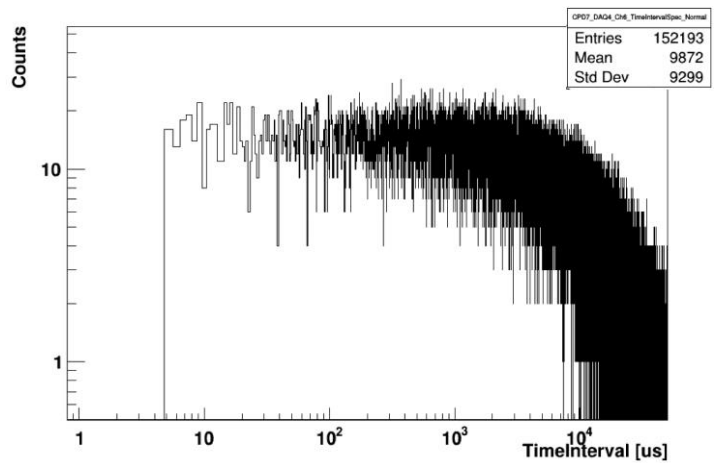

Figure 14 The time interval spectrum of CPD

# 4 Discussion and summary

CPD adopted a plastic scintillator-matched SiPM array for space charged particle detection. The capability of CPD to identify gamma-ray and charge particles combined with GRD was verified through simulation and experiment, based on the difference of detection efficiency of CPD and GRD for gamma rays and electrons.

Considering the temperature sensitivity of SiPM, we adopted the dual measures of temperature control and temperature gain compensation to solve the problem. In addition, there were full-component and semi-component operating modes for CPD based on reliability considerations. And the number of redundancies for CPD was designed to be three CPDs per satellite.

The performance indexes of the CPD are shown in Table 1. As a result, the energy range of electron, gamma-ray detection efficiency and dead time are tested to be better than the indexes required through the ground calibration experiment. And the energy response to the electron is consistent with the Geant4 simulation results.

Table 1 Performance indexes of CPD

| Item | Indexes required | Design/measured indexes |
| --- | --- | --- |
| Number | ≥5 | 8 |
| Plastic flash monomer size | ≥15 cm$^2$ | 16 cm$^2$ |
| energy range of electron | 300 keV–5 MeV | 210 keV–6.3 MeV |
| Gamma-ray efficiency | <20%@8–2000 keV | <14%@8–2000 keV |
| Dead time | ≤5 µs | 4.8 µs |

**Acknowledgements**

We would like to express our appreciation to the staff of Shandong Institute of aerospace electronic technology and Beijing Key Laboratory of Space Environment Exploration who offer great help in the phase of development. This research was supported by the "Strategic Priority Research Program" of the Chinese Academy of Sciences, Grant No. XDA 15360102. The authors also would thank the anonymous reviewers for their detailed and constructive comments in evaluation this paper.